\documentclass[twocolumn,secnumarabic,amssymb,amsmath,nofootinbib,tightenlines,showpacs,nobibnotes,aps,pra]{revtex4}
\usepackage{amsfonts}
\usepackage{bm}%
\usepackage{subfigure}%
\usepackage[colorlinks=true,linkcolor=blue]{hyperref}%
\usepackage{graphicx}% Include figure files
\usepackage{dcolumn}% Align table columns on decimal point

\begin{document}

\title{Fractional charge and statistics in the fractional quantum spin Hall effect}
\author{Yuanpei Lan}
\author{Shaolong Wan}
\altaffiliation{Corresponding author}
\email{slwan@ustc.edu.cn}
\affiliation{Institute for Theoretical
Physics and Department of Modern Physics University of Science and
Technology of China, Hefei, 230026, \textbf{P. R. China}}

\date{\today}

\begin{abstract}
In this paper, we consider there exist two types of fundamental
quasihole excitation in the fractional quantum spin Hall state and
investigate their topological properties by both Chern-Simons
field theory and Berry phase technique. By the two different ways,
we obtain the identical charge and statistical angle for each type
of quasihole, as well as the identical mutual statistics between
two different types of quasihole excitation.
\end{abstract}


\maketitle
%%%%%%%%%%%%%%%%%%%%%%%%%%%%%Main Body%%%%%%%%%%%%%%%%%%%%%%%%%%%%%%%%%%%%%

\section{Introduction}

In condensed matter systems, most states of matter can be
characterized by Landau's symmetry breaking theory. However,
topological states of quantum matter cannot be described by this
theory. The first examples of topological quantum states
discovered in nature are the quantum Hall (QH) states
\cite{Klitzing, Tsui} which opened up a new chapter in condensed
matter physics. In the noninteracting limit, the integer quantum
Hall (IQH) state is characterized by the TKNN invariant
\cite{Thouless} or first Chern number. For the fractional quantum
Hall (FQH) state, interaction between electrons turns out to be
crucial, and an Abelian FQH state can be characterized by an
integer symmetric $\mathbf{K}$ matrix and an integer charge vector
$\mathbf{t}$, up to a $SL(n,\mathbb{Z})$ equivalence \cite{Wen1}.
However, both the IQH and FQH states appear in a strong magnetic
field which breaks time reversal symmetry.

In a seminal paper, Haldane \cite{Haldane} has shown that the IQH
state can be realized in a tight-binding graphene lattice model
without net magnetic field, this state, however, breaks time
reversal symmetry due to a local magnetic flux density of a zero
net flux through the unit cell. Recently, Kane and Mele
\cite{Kane1, Kane2} proposed a novel class of topological state,
i.e. the integer quantum spin Hall (IQSH) state, which can be
viewed as two spin-dependent copies of Haldane's model and that
preserve time reversal symmetry. Bernevig and Zhang proposed the
QSH state for semiconductors \cite{Zhang1}, where the Landau
levels arise from the strain gradient, rather than the external
magnetic field. In the Landau level picture, the QSH state can be
understood as the state in which the spin-$\uparrow$ and
spin-$\downarrow$ electrons are in two opposing effective orbital
magnetic field $\mp B$ realized by spin-orbit (SO) coupling,
respectively. For the QSH state, the bulk is gapped and insulating
while there are gapless edge states in a time reversal invariant
system with SO coupling. Bernevig, Hughes and Zhang predict a QSH
state in HgTe/CdTe heterostructure \cite{Zhang2}, which has been
confirmed by experiment \cite{Zhang3}.

By analogy with the relation between the IQHE and FQHE, it is
natural to ask the question whether there can exist a fractional
QSH (FQSH) state. An explicit wave function for the FQSH state was
first proposed by Bernevig and Zhang \cite{Zhang1}, and the edge
theory was discussed by Levin \cite{Levin1}. Recently, a set of
exactly soluble lattice electron models for the FQSH state were
constructed \cite{Levin2}, and the generic wave function for the
FQSH state was proposed through Wannier function approach
\cite{Qi}.

As we all known that quasihole (or quasiparticle) excitations
above the FQH ground states have fractional charge and fractional
statistics \cite{Wilczek1,Nayak}, such as the $\nu$=$1/k$ Laughlin
states have quasihole excitations with charge $e/k$ and
statistical angle $\theta$=$\pi/k$. So we have to ask how the
properties of the excitations will be in FQSH state. In order to
answer this question, we investigate the charge and statistics of
the quasihole excitations in FQSH state. All five quantities for
the fractional charge and statistics are identically given by two
different ways. It is the main task in this paper.

The article is organized as follows. In Sec.\ref{sec2}, we
introduce the theoretical model of the QSHE and the wave function
for the FQSH state proposed by Bernevig and Zhang. In
Sec.\ref{sec3}, we first analyze the quasihole excitations in the
FQSH state and write down the wave function for each type of
fundamental quasihole excitation. And then the identical charge
and statistics of each type of excitation as well as the mutual
statistics between different types of excitations are obtained by
the two different method in Secs.\ref{sec3A} and \ref{sec3B},
respectively. Finally, Sec.\ref{sec4} is devoted to conclusions.

\section{Wave-function of fractional quantum spin hall state}
\label{sec2}

Now, we will brief review the theoretical model proposed by
Bernevig and Zhang \cite{Zhang1}. The simplest case of the QSHE
can be viewed as superposing two QH subsysyems with opposite
spins. The spin-$\uparrow$ QH state has positive charge hall
conduce ($\sigma_{xy} = +e^{2}/h$) while the spin-$\downarrow$ QH
state has negative charge Hall conductance ($\sigma_{xy} =
-e^{2}/h$). As such, the charge Hall conductance of the  whole
system vanishes. However, the spin Hall conductance remains finite
and quantized in units of $e/2\pi$ since the spin-$\uparrow$ and
the spin-$\downarrow$ QH states have opposite chirality.

To realize this QSH state we need a spin-dependent effective
orbital magnetic field, which can be created by the spin-orbit
coupling in conventional semiconductors in the presence of a
strain gradient. When the off-diagonal (shear) components of the
strain symmetric tensor are $\epsilon_{xy}(\leftrightarrow E_{z})
= 0$, $\epsilon_{xz}(\leftrightarrow E_{y}) = gy$,
$\epsilon_{yz}(\leftrightarrow E_{x}) = gx$, respectively, with g
denotes the magnitude of the strain gradient, a single electron in
a symmetric quantum well in the xy plane can be described by
\begin{eqnarray}
H=\frac{p_{x}^{2}+p_{y}^{2}}{2m}+\frac{1}{2}\frac{C_{3}}{\hbar}g(yp_{x}-xp_{y})\sigma_{z}+D(x^{2}+y^{2}),
\end{eqnarray}
where the second term corresponds to the spin-orbit coupling
$(\overrightarrow{p}\times\overrightarrow{E})\cdot\overrightarrow{\sigma}$,
and the third term is the confining potential. The constant
$C_{3}/\hbar$ is $8\times10^{5}$ m/s for GaAs. We introduce $l =
(8mD)^{-1/4}$ and $\omega = (8D/m)^{1/2}$, which have length
dimension and energy dimension separately, then the above
Hamiltonian can be expressed as
\begin{eqnarray}
H=\frac{1}{2}[p_{x}^2+p_{y}^2+\frac{1}{4}(x^{2}+y^{2})+R(yp_{x}-xp_{y})\sigma_{z}],
\end{eqnarray}
with $R$=$C_{3}g\sqrt{m/8D}$. For notation convenience, we use the
same notation for dimensionless $x(y)$, $p_{x}(p_{y})$ and $H$;
the correct units can be restored by introducing factors of $l$,
$l^{-1}$ and $\omega$. Since the eigenvalue of $\sigma_{z}$ is a
good quantum number, we can use the spin along the $z$ direction
to characterize the state and have
\begin{eqnarray}
 H&=&\left(
       \begin{array}{cc}
         H_{\uparrow} & 0 \\
         0 & H_{\downarrow} \\
       \end{array}
     \right)
     ,\nonumber\\
 H_{\downarrow,\uparrow}&=&\frac{1}{2}[p_{x}^2+p_{y}^2+(x^{2}+y^{2})\pm R(xp_{y}-yp_{x})].
\end{eqnarray}

We focus on the special case of $R=1$, i.e. $D = D_{0} =
mg^{2}C_{3}^{2}/8$, where the Hamiltonian becomes
\begin{eqnarray}
H&=&\frac{1}{2m}(\overrightarrow{p}+e\overrightarrow{A}\sigma_{z})^{2}, \nonumber\\
\overrightarrow{A}&=&\frac{mgC_{3}}{2e}(y,-x)=\frac{B}{2}(y,-x),
\end{eqnarray}
where $B=mgC_{3}/e$,
$\overrightarrow{B}=\nabla\times\overrightarrow{A}=-B\hat{z}$
plays the role of effective magnetic filed, which is essentially
arisen from the gradient of the strain. We notice, moreover, that
when $R=1$, the two quantities $l$ and $\omega$ become
$l_{0}=(8mD_{0})^{-1/4}=\sqrt{1/eB}$ and
$\omega=\omega_{0}=(8D_{0}/m)^{1/2}=eB/m$, which are exactly the
definitions of magnetic length $l_{B}$ and cyclotron frequency
$\omega_{c}$ in QH. Thus we can imagine the electrons of the
two-dimensional system experience a spin-dependent effective
magnetic field $-B\hat{z}\sigma_{z}$. We introduce the complex
coordinate $z= x+iy=re^{i\theta}$, $z^{*}=x-iy=re^{-i\theta}$, and
define the harmonic-oscillator ladder operators as
\begin{eqnarray}
a_{\uparrow}=\frac{1}{\sqrt{2}}(\frac{z}{2}+2\partial_{z^{*}}),~~~a_\uparrow^{\dagger}=\frac{1}{\sqrt{2}}(\frac{z^{*}}{2}-2\partial_{z}), \nonumber\\
a_{\downarrow}=\frac{1}{\sqrt{2}}(\frac{z^{*}}{2}+2\partial_{z}),~~~a_{\downarrow}^{\dagger}=\frac{1}{\sqrt{2}}(\frac{z}{2}-2\partial_{z{*}}).
\end{eqnarray}
They satisfy
$[a_{\sigma},a_{\sigma^{'}}^{\dagger}]=\delta_{\sigma,\sigma^{'}}$,
$[a_{\sigma},a_{\sigma^{'}}] = 0$,
($\sigma,\sigma^{'}=\{\uparrow,\downarrow$\}). In the vicinity of
$R=1$, the Hamiltonian can be written as
\begin{eqnarray}
H_{\uparrow}=a_{\uparrow}^{\dagger}a_{\uparrow}+\frac{1}{2},~~~~H_{\downarrow}=a_{\downarrow}^{\dagger}a_{\downarrow}+\frac{1}{2}.
\end{eqnarray}
We denote $m_{\uparrow}$ and $m_{\downarrow}$ as the eigenvalues
of $a_{\uparrow}^{\dagger}a_{\uparrow}$ and
$a_{\downarrow}^{\dagger}a_{\downarrow}$ respectively. The $z$
component of the angular momentum operator is given as
$L_{z}$=$-i\partial_{\theta}$=$z\partial_{z}-z^{*}\partial_{z^{*}}$=$a_{\downarrow}^{\dagger}a_{\downarrow}-a_\uparrow^{\dagger}a_\uparrow$,
which commutes with both $H_{\uparrow}$ and $H_{\downarrow}$, and
has eigenvalues $m_{\downarrow}-m_{\uparrow}$. For spin-$\uparrow$
electron, the lowest Landau level (LLL) corresponds to
$m_{\uparrow}$=$0$ and the single particle wave function is
$\psi_{m_{\downarrow}}^{\uparrow}$=$\frac{z^{m_\downarrow}}{\sqrt{2\pi2^{m_{\downarrow}}m_{\downarrow}!}}\exp(-\frac{1}{4}|z|^{2})$
with the angular momentum $m_{\downarrow}$. While for
spin-$\downarrow$ electrons, the LLL corresponds to
$m_{\downarrow}$=$0$ and the single particle wave function is
$\psi_{m_{\uparrow}}^{\downarrow}$=$\frac{(z^{*})^{m_\uparrow}}{\sqrt{2\pi2^{m_{\uparrow}}m_{\uparrow}!}}\exp(-\frac{1}{4}|z|^{2})$
with the angular momentum $-m_{\uparrow}$. It is easy to see that
the QSH system is equivalent to a bilayer system: in one layer we
have spin-$\uparrow$ electrons in the presence of a down-magnetic
filed($-B\hat{z}$), the spin-$\uparrow$ electrons are chiral and
have positive charge Hall conductance, while in other layer we
have spin-$\downarrow$ electrons in the presence of an up-magnetic
field($+B\hat{z}$), the spin-$\downarrow$ electrons are
anti-chiral and have negative charge Hall conductance.

In addition, the wave functions of spin-$\uparrow$ and
spin-$\downarrow$ are holomorphic and anti-holomorphic functions
respectively. If we consider the intra-layer correlations and
ignore inter-layer correlations, the many-body wave function would
be $\prod_{i<j}(z_{\uparrow i}-z_{\uparrow
j})^{m_{\uparrow}}\prod_{i<j}(z_{\downarrow i}^{*}-z_{\downarrow
j}^{*})^{m_{\downarrow}} e^{-1/4l_{B}^{2}(\sum_{i}|z_{\uparrow
i}|^{2}+\sum_{i}|z_{\downarrow i}|^{2})}$, where $z_{\uparrow i}$
and $z_{\downarrow i}$ denote the the spin-$\uparrow$ and
spin-$\downarrow$ coordinates respectively, and the filling
factors of the spin-$\uparrow$ and spin-$\downarrow$ layers are
$1/m_{\uparrow}$ and $1/m_{\downarrow}$ separately, the odd
integer $m_{\sigma}$ plays the role of the relative angular
momentum between electrons with spin orientation $\sigma$. Here,
$m_{\uparrow}$=$m_{\downarrow}$ because of the chiral-antichiral
symmetry.

However, the electrons in the QSH state reside in the same quantum
well and may possibly experience the additional interaction
between the spin-$\uparrow$ and spin-$\downarrow$ layers, then the
many-body wave function of the whole FQSH system should take the
form \cite{Zhang1}
\begin{eqnarray}
\Psi_{m_{\uparrow}m_{\downarrow}n}(\{z_{\uparrow i}\},\{z_{\downarrow i}^{*}\})
&=&\prod_{i<j}(z_{\uparrow i}-z_{\uparrow j})^{m_{\uparrow}}
\prod_{i<j}(z_{\downarrow i}^{*}-z_{\downarrow j}^{*})^{m_{\downarrow}}\nonumber\\
&&\times\prod_{i,j}(z_{\uparrow i}-z_{\downarrow j}^{*})^{n}\nonumber\\
&&\times e^{-\frac{1}{4l_{B}^{2}}(\sum_{i}|z_{\uparrow
i}|^{2}+\sum_{i}|z_{\downarrow i}|^{2})}. \label{wavefunction}
\end{eqnarray}

The above wave function is similar to the Halperin's wave function
\cite{Halperin}, which is a generalized Laughlin wave function to
a two-component system. The Halperin's wave functions are
holomorphic for both $z_{\uparrow i}$ and $z_{\downarrow i}$,
since for the bilayer QH system, the electrons of both layers
experience the same external magnetic filed. Whereas for the FQSH
system, the spin-$\uparrow$ electrons experience down effective
magnetic field and spin-$\downarrow$ electrons experience up
effective magnetic filed, this fact leads to the wave function of
FQSH incorporates both holomorphic and anti-holomorphic
coordinates.

\section{Fractional charge and fractional statistics in the FQSH system}
\label{sec3}

As we mentioned in Sec.\ref{sec2}, the FQSH system is analogous to
a bilayer FQH system, but the electrons in the FQSH system
experience spin-dependent effective magnetic field. For bilayer
FQH system, there are two kinds of fundamental quasihole
excitation which is in each layer, respectively \cite{Wen2}. Hence
it is natural to speculate that there exist two types of
fundamental quasihole excitations in the FQSH state. A quasihole
in the spin-$\uparrow$ layer is described by the wave function
\begin{eqnarray}
\Psi^{(\uparrow)}(\xi_{\uparrow})=\prod_{i}(\xi_{\uparrow}-z_{\uparrow
i})\Psi_{m_{\uparrow}m_{\downarrow}n}(\{z_{\uparrow
i}\},\{z_{\downarrow i}^{*}\}), \label{wavefunction up}
\end{eqnarray}
and in the spin-$\downarrow$ layer is described by
\begin{eqnarray}
\Psi^{(\downarrow)}(\xi_{\downarrow}^{*})=\prod_{i}(\xi_{\downarrow}^{*}-z_{\downarrow
i}^{*}) \Psi_{m_{\uparrow}m_{\downarrow}n}(\{z_{\uparrow
i}\},\{z_{\downarrow i}^{*}\}), \label{wavefunction down}
\end{eqnarray}
where $\xi_{\uparrow}$ and $\xi_{\downarrow}$ denote the
coordinates of quasiholes of the spin-$\uparrow$ and
spin-$\downarrow$ layers separately. In following, we calculate
the charge and statistics of the two types of the quasihole
excitation as well as the mutual statistics between a quasihole in
the spin-$\uparrow$ layer and a quasihole in the spin-$\downarrow$
layer, and give all five quantities for the fractional charge and
statistics by two different methods in Secs.\ref{sec3A} and
\ref{sec3B}, separately.

\subsection{Chern-Simons field theory approach}
\label{sec3A}

The topological structure of the FQH states can be understood by
several approaches and the most general is the low energy
effective Chern-Simons theory \cite{Wen1,Zhang4}. Wen and Zee
\cite{Wen1} pointed out that the Abelian FQH liquids can be
characterized by the integer valued $\mathbf{K}$-matrices and the
integer valued charged vectors, up to $SL(n,\mathbb{Z})$
equivalences, and the quasiparticle quantum numbers, such as
fractional charge and fractional statistics, can be calculated
from them. So we will search for the corresponding
$\mathbf{K}$-matrix and the charged vector $\mathbf{t}$ of the
FQSH state. Now we first construct the Chern-Simons theory for the
the FQSH state described by wave function (\ref{wavefunction}).

When $n$=$0$, the spin-$\uparrow$ electrons and spin-$\downarrow$
electrons decouple, the FQSH state is just two independent FQH
states with opposite magnetic fields. We introduce two U(1) gauge
fields $a_{\uparrow\mu}$, $a_{\downarrow\mu}$ to describe the
conserved electromagnetic current
$J_{\uparrow}^{\mu}=\frac{1}{2\pi}\varepsilon^{\mu \nu
\lambda}\partial_{\nu}a_{\uparrow\lambda}$ and
$J_{\downarrow}^{\mu}=\frac{1}{2\pi}\varepsilon^{\mu \nu
\lambda}\partial_{\nu}a_{\downarrow\lambda}$, separately. Coupling
the system to an external electromagnetic gauge potential
$A_{\mu}$ (here the notation $A_{\mu}$ is not the ``gauge
potential" of the effective orbital magnetic field in FQSH
system), then the total effective Lagrangian is
\begin{eqnarray}
\mathcal{L}=\mathcal{L}_{\uparrow}+\mathcal{L}_{\downarrow},
\end{eqnarray}
with
\begin{eqnarray}
\mathcal{L}_{\uparrow}=-\frac{m_{\uparrow}}{4\pi}\epsilon^{\mu\nu\lambda}a_{\uparrow\mu}\partial_{\nu}a_{\uparrow\lambda}
+\frac{e}{2\pi}\epsilon^{\mu\nu\lambda}A_{\mu}\partial_{\nu}a_{\uparrow\lambda}, \label{lagrangian up}\\
\mathcal{L}_{\downarrow}=+\frac{m_{\downarrow}}{4\pi}\epsilon^{\mu\nu\lambda}a_{\downarrow\mu}\partial_{\nu}a_{\downarrow\lambda}
+\frac{e}{2\pi}\epsilon^{\mu\nu\lambda}A_{\mu}\partial_{\nu}a_{\downarrow\lambda}.
\label{lagrangian down}
\end{eqnarray}

When we integrate out $a_{\uparrow}$ and $a_{\downarrow}$ from
Eqs.(\ref{lagrangian up}) and (\ref{lagrangian down}), one can
obtain the linear response of the both layers to the external
electromagnetic fields. In order to correctly reflect the fact
that the spin-$\uparrow$ FQH layer has positive charge Hall
conductance while the spin-$\downarrow$ FQH layer has negative
charge Hall conductance, we must use the negative sign in front of
$a_{\uparrow}\wedge{da_{\uparrow}}$ and the positive sign in front
of $a_{\downarrow}\wedge{da_{\downarrow}}$.

We now turn to the case for $n\neq0$, i.e. there exist the
inter-layer correlations between the spin-$\uparrow$ electrons and
spin-$\downarrow$ electrons. We start with the spin-$\downarrow$
FQH layer, namely $\prod_{i<j}(z_{\downarrow i}^{*}-z_{\downarrow
j}^{*})^{m_{\downarrow}}\exp(-\frac{1}{4l_{B}^{2}}\sum_{i}|z_{\downarrow
i}|^{2})$, which is a $1/m_{\downarrow}$ Laughlin state and can be
described by the effective Lagrangian (\ref{lagrangian down}).
Since the factor $\prod_{i}(z_{\uparrow i}-\xi_{\uparrow})$ in
(\ref{wavefunction up}) corresponds to a fundamental quasihole
excitation in the spin-$\uparrow$ layer, from the factor
$\prod_{i,j}(z_{\uparrow i}-z_{\downarrow j}^{*})^{n}$ in
(\ref{wavefunction}) we can conclude that an electron in the
spin-$\uparrow$ layer is bound to a ``large" quasihole excitation
in the spin-$\downarrow$ layer. Such a quasihole is n-fold
fundamental quasihole and carries an $a_{\downarrow\mu}$ charge of
$-n$. While in the mean filed theory (MFT), we view the
$a_{\downarrow\mu}$ field as a fixed background and ignore the
response of the $a_{\downarrow\mu}$ field, then the quasihole gas
behaves like bosons in the ``magnetic" filed
$nb_{\downarrow}=-n\varepsilon_{ij}\partial_{i}a_{\downarrow j}$.
We can introduce a bosonic field $\phi$ to describe the quasihole
excitations by
\begin{eqnarray}
\phi^{\dagger}i(\partial_{0}+ina_{\downarrow
0})\phi+\frac{1}{2m}\phi^{\dagger}(\partial_{i}+ina_{\downarrow
i})^{2}\phi.
\end{eqnarray}

When we attach a spin-$\uparrow$ electron to each quasihole (i.e.
bosonic field $\phi$ in the MFT), the bound state behaves like a
fermion, we denote it by $\psi$ field, these fermions experience
an effective magnetic field $-eB+nb_{\downarrow}$, this process
can be describe by
\begin{eqnarray}
\psi^{\dagger}i(\partial_{0}-ieA_{0}+ina_{\downarrow
0})\psi+\frac{1}{2m}\psi^{\dagger}(\partial_{i}-ieA_{i}+ina_{\downarrow
i})^{2}\psi. \nonumber\\ &&
\end{eqnarray}
When the fermion density satisfies
\begin{eqnarray}
\psi^{\dagger}\psi=\frac{1}{m_{\uparrow}}\frac{-eB+nb_{\downarrow}}{2\pi},
\nonumber
\end{eqnarray}
(the effective orbital magnetic field is negative for
spin-$\uparrow$ electron as mentioned in section \ref{sec2}), i.e.
the fermions have filling factor $\frac{1}{m_{\uparrow}}$, the
ground state of the electrons in the spin-$\uparrow$ layer
condensate to a $\frac{1}{m_{\uparrow}}$ Laughlin state, which
corresponds to the factor $\prod_{i<j}(z_{\uparrow i}-z_{\uparrow
j})^{m_{\uparrow}}$ in Eq.(\ref{wavefunction}). And then we
introduce another gauge field $a_{\uparrow \mu}$ to describe the
conversed fermion current $j^{\mu}=\frac{1}{2\pi}\varepsilon^{\mu
\nu \lambda}\partial_{\nu}a_{\uparrow\lambda}$ in the
spin-$\uparrow$ layer, the effective theory of
$\frac{1}{m_{\uparrow}}$ Laughlin state in the spin-$\uparrow$
layer is given by
\begin{eqnarray}
\mathcal{L}_{\uparrow}^{'}=-\frac{m_{\uparrow}}{4\pi}\epsilon^{\mu\nu\lambda}a_{\uparrow\mu}\partial_{\nu}a_{\uparrow\lambda}.
\label{lagrangian up'}
\end{eqnarray}
The effective theory of the bound states has a form
\begin{eqnarray}
\mathcal{L}_{b}&=&(eA_{\mu}-na_{\downarrow \mu})j^{\mu}\nonumber\\
&=&\frac{e}{2\pi}\epsilon^{\mu\nu\lambda}A_{\mu}\partial_{\nu}a_{\uparrow\lambda}
-\frac{n}{2\pi}\epsilon^{\mu\nu\lambda}a_{\downarrow\lambda}\partial_{\nu}a_{\uparrow\lambda}.
\label{lagrangian b}
\end{eqnarray}

Putting (\ref{lagrangian down}), (\ref{lagrangian up'}) and
(\ref{lagrangian b}) together, we eventually obtain the
Chern-Simons theory of the FQSH state, i.e.
\begin{eqnarray}
\mathcal{L}=-\frac{1}{4\pi}\sum_{I,J}a_{I\mu}K_{IJ}\epsilon^{\mu\nu\lambda}\partial_{\nu}a_{J\lambda}
+\frac{e}{2\pi}\sum_{I}t_{I}A_{\mu}\epsilon^{\mu\nu\lambda}\partial_{\nu}a_{I\lambda}, \nonumber\\
&& \label{lagrangian FQSH G}
\end{eqnarray}
where $I,J=\{\uparrow,\downarrow\}$ and
\begin{eqnarray}
\mathbf{K}=\left(
    \begin{array}{cc}
      m_{\uparrow} & n \\
      n & -m_{\downarrow} \\
    \end{array}
  \right),~~~~
  \mathbf{t}=\left(
                       \begin{array}{c}
                         1 \\
                         1 \\
                       \end{array}
                     \right). \label{K,t}
\end{eqnarray}

A generic quasiparticle (or quasihole) in the FQSH state is a
bound state which carries $l_{\uparrow}$ units of $a_{\uparrow
\mu}$ charge and $l_{\downarrow}$ units of $a_{\downarrow \mu}$
charge, we can denote it by a vector
$\mathbf{l}$=$(l_{\uparrow},l_{\downarrow})^{T}$. If there are k
quasiparticles in the system, each of them is labelled as
$\mathbf{l}^{(L)}$=$(l_{\uparrow}^{(L)},l_{\downarrow}^{(L)})^{T}$,
$(L=1,2,\cdots,k)$, moreover, we use $j_{L}$ denote the current of
the Lth quasiparticle, then the effective Lagrangian of these
quasiparticles has a form
\begin{eqnarray}
\Delta\mathcal{L}=\sum_{I}a_{I\mu}j_{I}^{\mu},~~~I=\{\uparrow,\downarrow\},
\label{lagrangian delta}
\end{eqnarray}
where
\begin{eqnarray}
j_{I}^{\mu}=l_{I}^{(1)}j_{1}^{\mu}+l_{I}^{(2)}j_{2}^{\mu}+\cdots+l_{I}^{(k)}j_{k}^{\mu}=\sum_{L=1}^{k}l_{I}^{(L)}j_{L}^{\mu}.
\end{eqnarray}

So the total low energy effective theory of the FQSH state with
quasiparticle excitations is the sum of (\ref{lagrangian FQSH G})
and (\ref{lagrangian delta}):
\begin{eqnarray}
\mathcal{L}_{t}&=&-\frac{1}{4\pi}\sum_{I,J}a_{I\mu}K_{IJ}\epsilon^{\mu\nu\lambda}\partial_{\nu}a_{J\lambda}\nonumber\\
&&+\frac{e}{2\pi}\sum_{I}t_{I}A_{\mu}\epsilon^{\mu\nu\lambda}\partial_{\nu}a_{I\lambda}+\sum_{I}a_{I\mu}j_{I}^{\mu}.
\label{lagrangian FQSH T}
\end{eqnarray}
Integrating out the gauge filed $a_{I}$ entirely,
\begin{eqnarray}
e^{i\int d^{3}x\mathcal{L}_{Hopf}}=\int\mathcal{D}a e^{i\int
d^{3}x\mathcal{L}_{t}}, \nonumber
\end{eqnarray}
we obtain a matrix version of the nonlocal Hopf Lagrangian \cite{Wilczek2}
\begin{eqnarray}
\mathcal{L}_{Hopf}=\pi\tilde{j}_{I}^{\mu}K_{IJ}^{-1}(\frac{\epsilon_{\mu\nu\lambda}\partial^{\nu}}{\partial^{2}})\tilde{j}_{J}^{\lambda},
\label{lagrangian Hopf}
\end{eqnarray}
with the modified currents
\begin{eqnarray}
\tilde{j}_{I}^{\mu}={j}_{I}^{\mu}+\frac{e}{2\pi}t_{I}\epsilon^{\mu\nu\lambda}\partial_{\nu}A_{\lambda}.
\end{eqnarray}

The Lagrangian (\ref{lagrangian Hopf}) contains three types of
terms: $AA$, $Aj$ and $jj$. The $Aj$ term determine the electronic
charge of the Lth quasiparticle:
\begin{eqnarray}
Q^{(L)}/e=\sum_{I,J}t_{I}K_{IJ}^{-1}l_{J}^{(L)}=\mathbf{t}^{T}\mathbf{K}^{-1}\mathbf{l}.
\label{Eq charge}
\end{eqnarray}
The quasiparticles interact with each other via the $jj$ term, and
the value of $S$=$\pi j
K^{-1}\frac{\varepsilon\partial}{\partial^{2}}j$ can be determined
by relating the Hopf invariant to the Gauss's linking number
between the trajectories of two quasiparticles \cite{Polyakov}. So
when we move the $L$th quasiparticle all the way around the
$L^{'}$th quasiparticle the wave function acquires a phase angle
$\theta^{(LL^{'})}$:
\begin{eqnarray}
\frac{\theta^{(LL^{'})}}{2\pi}=\sum_{I,J}l_{I}K_{IJ}^{-1}l_{J}^{'}=\mathbf{l}^{T}\mathbf{K}^{-1}\mathbf{l}^{'}.
\label{Eq mutual statistics}
\end{eqnarray}
For $L$ = $L^{'}$, the statistical angle
$\theta^{(L)}$=$\theta^{(LL)}/2$ associated with exchanging two
identical particles is given by:
\begin{eqnarray}
\frac{\theta^{(L)}}{\pi}=\sum_{I,J}l_{I}K_{IJ}^{-1}l_{J}=\mathbf{l}^{T}\mathbf{K}^{-1}\mathbf{l}.
\label{Eq statistical angle}
\end{eqnarray}
The two types of fundamental quasihole excitation in the FQSH
state are labelled by
\begin{eqnarray}
\mathbf{l}_{\uparrow}=\left(
                        \begin{array}{c}
                          1 \\
                          0 \\
                        \end{array}
                      \right)
                      ~,~~~
\mathbf{l}_{\downarrow}=\left(
                        \begin{array}{c}
                          0 \\
                          1 \\
                        \end{array}
                      \right). \label{two quasihole}
\end{eqnarray}

Then, according to Eqs.(\ref{Eq charge}) and (\ref{two
quasihole}), the physical electronic charge of the fundamental
quasihole excitations of the spin-$\uparrow$ and spin-$\downarrow$
layers are given as
\begin{eqnarray}
Q^{(\uparrow)}=e\frac{m_{\downarrow}+n}{m_{\uparrow}m_{\downarrow}+n^{2}}~,~~~Q^{(\downarrow)}=e\frac{-m_{\uparrow}+n}{m_{\uparrow}m_{\downarrow}+n^{2}}.
\label{quasihole charge}
\end{eqnarray}

According to Eqs.(\ref{Eq statistical angle}) and (\ref{two
quasihole}), the statistical angles for the two types of the
fundamental quasihole excitations are given as
\begin{eqnarray}
\theta^{(\uparrow)}=\pi\frac{m_{\downarrow}}{m_{\uparrow}m_{\downarrow}+n^{2}}~,
~~~\theta^{(\downarrow)}=-\pi\frac{m_{\uparrow}}{m_{\uparrow}m_{\downarrow}+n^{2}}.
\label{quasihole statistical angles}
\end{eqnarray}

Applying Eqs.(\ref{Eq mutual statistics}) and (\ref{two
quasihole}), we can get the mutual statistics between a quasihole
in the spin-$\uparrow$ lay and a quasihole in the
spin-$\downarrow$ layer:
\begin{eqnarray}
\theta^{(\uparrow\downarrow)}=2\pi\frac{n}{m_{\uparrow}m_{\downarrow}+n^{2}}.
\label{quasiholes mutual statistics}
\end{eqnarray}

\subsection{Berry phase technique}
\label{sec3B}

The fractional charge and fractional statistics of the excitations
of the $\nu$=$1/k$ Laughlin FQH state can be calculated directly
by Berry phase, which is a simple but profound concept relating to
the adiabatic theorem in quantum mechanics. The basic idea is as
follows \cite{Arovas1}: We compute the Berry phase under the
assumption that the quasihole is slowly transported around a loop.
The calculation will enable us to infer the charge of the
quasihole, while the statistical angle can be obtained by
computing the Berry phase when one quasihole winds around another.
Now we will use the similar idea to calculate the fractional
statistics in the FQSH state.

As we mentioned in Sec.\ref{sec2}, the wave function of
spin-$\uparrow$ (-$\downarrow$) electrons are holomorphic
(antiholomorphic), which is caused by the opposite direction of
the effective magnetic field for two layers. In other words, if we
change the direction of the effective magnetic field in the
spin-$\downarrow$ layer, then the wave function of
spin-$\downarrow$ electrons becomes holomorphic, meanwhile, the
``effective" filling factor of spin-$\downarrow$ electrons becomes
a negative number. In this sense, we consider there exist a
one-to-one correspondence between the wave function
(\ref{wavefunction}) and the following wave function
\begin{widetext}
\begin{eqnarray}
\Psi_{m_{\uparrow}-m_{\downarrow}n}(\{z_{\uparrow
i}\},\{z_{\downarrow i}\}) &=&\prod_{i<j}(z_{\uparrow
i}-z_{\uparrow j})^{m_{\uparrow}} \prod_{i<j}(z_{\downarrow
i}-z_{\downarrow j})^{-m_{\downarrow}} \prod_{i,j}(z_{\uparrow
i}-z_{\downarrow j})^{n}
e^{-\frac{1}{4l_{B}^{2}}(\sum_{i}|z_{\uparrow
i}|^{2}+\sum_{i}|z_{\downarrow i}|^{2})}. \label{wavefunction '}
\end{eqnarray}
\end{widetext}

The corresponding wave functions of the quasihole excitations in
the spin-$\uparrow$ and the spin-$\downarrow$ layers are
\begin{eqnarray}
\Psi^{(\uparrow)}(\xi_{\uparrow})=\prod_{i}(\xi_{\uparrow}-z_{\uparrow
i})\Psi_{m_{\uparrow}-m_{\downarrow}n}(\{z_{\uparrow
i}\},\{z_{\downarrow i}\}),
\label{wavefunction up '}\\
\Psi^{(\downarrow)}(\xi_{\downarrow})=\prod_{i}(\xi_{\downarrow}-z_{\downarrow
i})\Psi_{m_{\uparrow}-m_{\downarrow}n}(\{z_{\uparrow
i}\},\{z_{\downarrow i}\}). \label{wavefunction down'}
\end{eqnarray}

The plasma analogy \cite{Laughlin} results from writing the
quantum distribution function, the square of the many body wave
function, as a classical statistical mechanics distribution
function for interacting particles in an external potential, i.e.
\begin{eqnarray}
|\Psi|^{2}=e^{-U}. \label{plasma}
\end{eqnarray}
The classical system described by Eq.(\ref{plasma}) is a
two-dimensional generalized Coulomb plasma \cite{Forrester}, in
which there exist two species of particles and interactions
between them are logarithmic potentials with three independent
coupling constants. According to Eqs.(\ref{wavefunction up '}) and
(\ref{plasma}), the classical potential energy corresponding to
$\Psi^{(\uparrow)}$ is
\begin{widetext}
\begin{eqnarray}
U^{(\uparrow)}&=&m_{\uparrow}\sum_{i<j}(-2\ln|z_{\uparrow
i}-z_{\uparrow
j}|)+(-m_{\downarrow})\sum_{i<j}(-2\ln|z_{\downarrow
i}-z_{\downarrow j}|)
+n\sum_{i<j}(-2\ln|z_{\uparrow i}-z_{\downarrow j}|)\nonumber\\
&&+\sum_{i}\frac{|z_{\uparrow
i}|^{2}}{2l_{B}^{2}}+\sum_{i}\frac{|z_{\downarrow
i}|^{2}}{2l_{B}^{2}} +\sum_{i}(-2\ln|\xi_{\uparrow}-z_{\uparrow
i}|). \label{Eq potential}
\end{eqnarray}
\end{widetext}
This potential energy function (\ref{Eq potential}) describes a
system consisting of pseudospin-$\uparrow$ and
pseudospin-$\downarrow$ particles (Here we only apply $\uparrow$
and $\downarrow$ to denote the two species of particles in
two-dimensional plasma, the particle index is called
``pseudospin".). All particles have mutual two-dimensional Coulomb
interactions with coupling constant $m_{\uparrow}$ between two
pseudospin-$\uparrow$ particles, coupling constant
$-m_{\downarrow}$ between two pseudospin-$\downarrow$ particles,
and coupling constant $n$ between a pseudospin-$\uparrow$ particle
and pseudospin-$\downarrow$ particle. Meanwhile, both the
pseudospin-up and pesudospin-down particles interact with  unit
coupling constant with the uniform background, and the uniform
background has charge density $(2\pi l_{B}^{2})^{-1}$. From
Eq.(\ref{Eq potential}) we know that for $U^{(\uparrow)}$, only
pseudospin-$\uparrow$ particles interact with a unit coupling
constant with an pseudospin-$\uparrow$ impurity particle located
at the $\xi_{\uparrow}$, while pseudospin-$\downarrow$ particles
do not interact with an pseudospin-$\uparrow$ impurity particle at
all. The charge densities induced in each species of particle by
the impurity can be calculated using the perfect screening
properties that result from the long range interactions of the
plasma. Far enough from the impurity the net interaction must
vanish for each species of particle. In other words, the sum of
the impurity charge times its coupling constant plus the induced
charge in each plasma component times the coupling strength for
that plasma component must vanish, i.e.
\begin{eqnarray}
\left(
  \begin{array}{cc}
    m_{\uparrow} & n \\
    n & -m_{\downarrow} \\
  \end{array}
\right)\left(
         \begin{array}{c}
           q_{\uparrow}^{(\uparrow)} \\
           q_{\downarrow}^{(\uparrow)} \\
         \end{array}
       \right)=\left(
                 \begin{array}{c}
                   -1 \\
                   0 \\
                 \end{array}
               \right), \label{Eq charge up up  down}
\end{eqnarray}
where $q_{\uparrow}^{(\uparrow)}$ and
$q_{\downarrow}^{(\uparrow)}$ represent the pesudospin-$\uparrow$
component and the pesudospin-$\downarrow$ component of the induced
charge by the pseudospin-$\uparrow$ impurity. From Eq.(\ref{Eq
charge up up  down}), we get
\begin{eqnarray}
q_{\uparrow}^{(\uparrow)}=-\frac{m_{\downarrow}}{m_{\uparrow}m_{\downarrow}+n^{2}}~,
~~q_{\downarrow}^{(\uparrow)}=-\frac{n}{m_{\uparrow}m_{\downarrow}+n^{2}}.
\end{eqnarray}
And the total physical electronic charge of the quasihole in the spin-$\uparrow$ layer is
\begin{eqnarray}
Q^{(\uparrow)}=-e(q_{\uparrow}^{(\uparrow)}+q_{\downarrow}^{(\uparrow)})=e\frac{m_{\downarrow}+n}{m_{\uparrow}m_{\downarrow}+n^{2}}.
\label{Eq charge up B}
\end{eqnarray}
Similarly, for wave function $\Psi^{(\downarrow)}$, we get
\begin{eqnarray}
\left(
  \begin{array}{cc}
    m_{\uparrow} & n \\
    n & -m_{\downarrow} \\
  \end{array}
\right)\left(
         \begin{array}{c}
           q_{\uparrow}^{(\downarrow)} \\
           q_{\downarrow}^{(\downarrow)} \\
         \end{array}
       \right)=\left(
                 \begin{array}{c}
                   0 \\
                   -1 \\
                 \end{array}
               \right), \label{Eq charge down up down}
\end{eqnarray}
where $q_{\uparrow}^{(\downarrow)}$ and
$q_{\downarrow}^{(\downarrow)}$ represent the
pseudospin-$\uparrow$ component and the pseudospin-$\downarrow$
component of the induced charge by the pseudospin-$\downarrow$
impurity. From Eq.(\ref{Eq charge down up down}), we get
\begin{eqnarray}
q_{\uparrow}^{(\downarrow)}=-\frac{n}{m_{\uparrow}m_{\downarrow}+n^{2}}~,
~~q_{\downarrow}^{(\downarrow)}=\frac{m_{\uparrow}}{m_{\uparrow}m_{\downarrow}+n^{2}}.
\end{eqnarray}
And the total physical electronic charge of the quasihole in the spin-$\downarrow$ layer is
\begin{eqnarray}
Q^{(\downarrow)}=-e(q_{\uparrow}^{(\downarrow)}+q_{\downarrow}^{(\downarrow)})=e\frac{-m_{\uparrow}+n}{m_{\uparrow}m_{\downarrow}+n^{2}}.
\label{Eq charge down B}
\end{eqnarray}
We find that Eqs.(\ref{Eq charge up B}) and (\ref{Eq charge down
B}) are in perfect agreement with the results (\ref{quasihole
charge}) in Sec.\ref{sec3A}.

The normalized wave function corresponding to (\ref{wavefunction
up '}) is given by
\begin{eqnarray}
\Psi^{(\uparrow)}(\xi_{\uparrow},\xi_{\uparrow}^{*})=\frac{1}{\sqrt{N^{(\uparrow)}(\xi_{\uparrow},\xi_{\uparrow}^{*})}}\Psi^{(\uparrow)}(\xi_{\uparrow}),
\end{eqnarray}
where
$N^{(\uparrow)}(\xi_{\uparrow},\xi_{\uparrow}^{*})$=$\langle\Psi^{(\uparrow)}(\xi_{\uparrow})|\Psi^{(\uparrow)}(\xi_{\uparrow})\rangle$
is the normalization factor. Assuming that a quasihole is slowly
transported around a loop $\Gamma$, the phase change of
$\Psi^{(\uparrow)}(\xi_{\uparrow}(t),\xi_{\uparrow}^{*}(t))$ gives
the Berry phase $e^{i\gamma}$,
\begin{eqnarray}
\gamma&=&i\int_{t_{0}}^{t_{1}}dt
\langle\Psi^{(\uparrow)}(\xi_{\uparrow}(t),\xi_{\uparrow}^{*}(t))|\frac{d}{dt}|\Psi^{(\uparrow)}(\xi_{\uparrow}(t),\xi_{\uparrow}^{*}(t))\rangle\nonumber\\
&=&\oint(a_{\xi_{\uparrow}}d\xi_{\uparrow}+a_{\xi_{\uparrow}^{*}}d\xi_{\uparrow}^{*}),
\label{Eq Berry phase}
\end{eqnarray}
with
\begin{eqnarray}
a_{\xi_{\uparrow}}=
i\langle\Psi^{(\uparrow)}(\xi_{\uparrow},\xi_{\uparrow}^{*})|\frac{\partial}{\partial\xi_{\uparrow}}|\Psi^{(\uparrow)}(\xi_{\uparrow},\xi_{\uparrow}^{*})\rangle,
\nonumber\\
a_{\xi_{\uparrow}^{*}}=
i\langle\Psi^{(\uparrow)}(\xi_{\uparrow},\xi_{\uparrow}^{*})|\frac{\partial}{\partial\xi_{\uparrow}^{*}}|\Psi^{(\uparrow)}(\xi_{\uparrow},\xi_{\uparrow}^{*})\rangle,
\end{eqnarray}
moreover $a_{\xi_{\uparrow}}$ and
$a_{\xi_{\uparrow}^{*}}$ can be expressed in terms of the
normalization factor:
\begin{eqnarray}
a_{\xi_{\uparrow}}=\frac{i}{2}\frac{\partial}{\partial\xi_{\uparrow}}\ln
N^{(\uparrow)},~~
a_{\xi_{\uparrow}^{*}}=-\frac{i}{2}\frac{\partial}{\partial\xi_{\uparrow}^{*}}\ln
N^{(\uparrow)}. \label{Eq. a, a*}
\end{eqnarray}
From Eqs.(\ref{Eq Berry phase}) and (\ref{Eq. a, a*}), we know
that in order to get the Berry phase, we should calculate the
normalization factor which can be obtained from the generalized
plasma analogue. Because of the complete screening of the plasma,
the net force acting on the volume element of the background
vanishes, i.e.
$q_{\uparrow}^{(\uparrow)}\cdot1+q_{\downarrow}^{(\uparrow)}\cdot1+1\cdot
g^{(\uparrow b)}=0$, with $g^{(\uparrow b)}$ denotes the coupling
constant between the pseudospin-$\uparrow$ impurity and the
background charge, then we have
\begin{eqnarray}
g^{(\uparrow
b)}=-(q_{\uparrow}^{(\uparrow)}+q_{\downarrow}^{(\uparrow)})
=\frac{m_{\downarrow}+n}{m_{\uparrow}m_{\downarrow}+n^{2}}.
\end{eqnarray}

Let us consider a state characterized by the presence of two
quasihole excitations in the spin-$\uparrow$ layer. Assuming that
the quasihole in $\xi_{\uparrow}$ is adiabatically moved on a full
loop $\Gamma$, while the quasihole in $\xi_{\uparrow}^{'}$ is
fixed. Again, due to the complete screening of the plasma, the net
force on the quasihole in $\xi_{\uparrow}^{'}$ vanishes, i.e.
$q_{\uparrow}^{(\uparrow)}\cdot1+q_{\downarrow}^{(\uparrow)}\cdot0+1\cdot
g^{(\uparrow\uparrow)}=0$, with $g^{(\uparrow\uparrow )}$ denotes
the coupling constant between the two pseudospin-$\uparrow$
impurities, then we have
\begin{eqnarray}
g^{(\uparrow\uparrow
)}=-q_{\uparrow}^{(\uparrow)}=\frac{m_{\downarrow}}{m_{\uparrow}m_{\downarrow}+n^{2}}.
\end{eqnarray}
We note that, after including a term
$\exp(-\frac{m_{\downarrow}+n}{m_{\uparrow}m_{\downarrow}+n^{2}}\frac{|\xi_{\uparrow}|^{2}}{4l_{B}^{2}})$
corresponding to the interaction between the pseudospin-$\uparrow$
impurity and the background $g^{(\uparrow
b)}\frac{|\xi_{\uparrow}|^{2}}{2l_{B}^{2}}$, and a term
$|\xi_{\uparrow}-\xi_{\uparrow}^{'}|^{\frac{m_{\downarrow}}{m_{\uparrow}m_{\downarrow}+n^{2}}}$
corresponding to the interaction between the two
pseudospin-$\uparrow$ impurities
$g^{(\uparrow\uparrow)}(-2\ln|\xi_{\uparrow}-\xi_{\uparrow}^{'}|)$,
the total energy
$U(\xi_{\uparrow},\xi_{\uparrow}^{*},\xi_{\uparrow}^{'},\xi_{\uparrow}^{'*})$
of the plasma with two pseudospin-$\uparrow$ impurities is given
by:
\begin{eqnarray}
e^{-U(\xi_{\uparrow},\xi_{\uparrow}^{*},\xi_{\uparrow}^{'},\xi_{\uparrow}^{'*})}&=&
|e^{-\frac{m_{\downarrow}+n}{m_{\uparrow}m_{\downarrow}+n^{2}}\frac{1}{4l_{B}^{2}}(|\xi_{\uparrow}|^{2}+|\xi_{\uparrow}^{'}|^{2})}
\nonumber\\
&&\times|\xi_{\uparrow}-\xi_{\uparrow}^{'}|^{\frac{m_{\downarrow}}{m_{\uparrow}m_{\downarrow}+n^{2}}}\nonumber\\
&&\times\Psi^{(\uparrow\uparrow)}(\xi_{\uparrow},\xi_{\uparrow}^{'})|^{2},
\label{Eq energy1}
\end{eqnarray}
where
\begin{eqnarray}
\Psi^{(\uparrow\uparrow)}(\xi_{\uparrow},\xi_{\uparrow}^{'})
&=&\prod_{i}(\xi_{\uparrow}-z_{\uparrow i})\prod_{i}(\xi_{\uparrow}^{'}-z_{\uparrow i})\nonumber\\
&&\times\Psi_{m_{\uparrow}-m_{\downarrow}n}(\{z_{\uparrow
i}\},\{z_{\downarrow i}\}), \label{Eq nenrgy2}
\end{eqnarray}
and the corresponding normalization factor is
\begin{eqnarray}
N^{(\uparrow\uparrow)}(\xi_{\uparrow},\xi_{\uparrow}^{*},\xi_{\uparrow}^{'},\xi_{\uparrow}^{'*})
=\langle\Psi^{(\uparrow\uparrow)}(\xi_{\uparrow},\xi_{\uparrow}^{'})|\Psi^{(\uparrow\uparrow)}(\xi_{\uparrow},\xi_{\uparrow}^{'})\rangle.
\label{Eq energy3}
\end{eqnarray}
Due to the complete screening of the plasma, the total energy
$U(\xi_{\uparrow},\xi_{\uparrow}^{*},\xi_{\uparrow}^{'},\xi_{\uparrow}^{'*})$
of the plasma is independent of $\xi_{\uparrow}$($\xi_{\uparrow}^{*}$) and $\xi_{\uparrow}$($\xi_{\uparrow}^{'*}$), i.e.
\begin{eqnarray}
U(\xi_{\uparrow},\xi_{\uparrow}^{*},\xi_{\uparrow}^{'},\xi_{\uparrow}^{'*})=const. \label{Eq energy4}
\end{eqnarray}
From Eqs.(\ref{Eq energy1})-(\ref{Eq energy4}), we find that
\begin{eqnarray}
N^{(\uparrow\uparrow)}(\xi_{\uparrow},\xi_{\uparrow}^{*},\xi_{\uparrow}^{'},\xi_{\uparrow}^{'*})&\propto&
e^{\frac{m_{\downarrow}+n}{m_{\uparrow}m_{\downarrow}+n^{2}}\frac{1}{2l_{B}^{2}}(|\xi_{\uparrow}|^{2}+|\xi_{\uparrow}^{'}|^{2})}\nonumber\\
&&\times
|\xi_{\uparrow}-\xi_{\uparrow}^{'}|^{-\frac{2m_{\downarrow}}{m_{\uparrow}m_{\downarrow}+n^{2}}}.
\label{Eq N}
\end{eqnarray}
Without loss of generality, setting $\xi_{\uparrow}^{'}$=$0$ ,
combining Eqs.(\ref{Eq. a, a*}) and (\ref{Eq N}), we have
\begin{eqnarray}
a_{\xi_{\uparrow}}&=&i\frac{m_{\downarrow}+n}{m_{\uparrow}m_{\downarrow}+n^{2}}\frac{1}{4l_{B}^{2}}\xi_{\uparrow}^{*}
-\frac{i}{2}\frac{m_{\downarrow}}{m_{\uparrow}m_{\downarrow}+n^{2}}\frac{1}{\xi_{\uparrow}}, \nonumber\\
a_{\xi_{\uparrow}^{*}}&=&-i\frac{m_{\downarrow}+n}{m_{\uparrow}m_{\downarrow}+n^{2}}\frac{1}{4l_{B}^{2}}\xi_{\uparrow}
+\frac{i}{2}\frac{m_{\downarrow}}{m_{\uparrow}m_{\downarrow}+n^{2}}\frac{1}{\xi_{\uparrow}^{*}}.
\label{Eq a, a* E}
\end{eqnarray}
From Eqs.(\ref{Eq Berry phase}) and (\ref{Eq a, a* E}), we have
\begin{eqnarray}
\gamma&=&i\frac{m_{\downarrow}+n}{m_{\uparrow}m_{\downarrow}+n^{2}}\frac{1}{4l_{B}^{2}}(\oint\xi_{\uparrow}^{*}d\xi_{\uparrow}-\oint\xi_{\uparrow}
d\xi_{\uparrow}^{*})\nonumber\\
&&-\frac{i}{2}\frac{m_{\downarrow}}{m_{\uparrow}m_{\downarrow}+n^{2}}(\oint\frac{1}{\xi_{\uparrow}}d\xi_{\uparrow}-\oint\frac{1}{\xi_{\uparrow}^{*}}d\xi_{\uparrow}^{*}).
\end{eqnarray}
As we move the quasihole $\xi_{\uparrow}$ around $\xi_{\uparrow}^{'}$=$0$ a circle of radius $r$(=const.), i.e. $\xi_{\uparrow}$=$re^{i\theta}$
then $\xi_{\uparrow}^{*}$=$r^{2}(\xi_{\uparrow})^{-1}$,
$\oint d\xi_{\uparrow}/\xi_{\uparrow}$=$-\oint d\xi_{\uparrow}^{*}/\xi_{\uparrow}^{*}$=$2\pi i$, we have
\begin{eqnarray}
\gamma=-\pi
r^{2}\frac{m_{\downarrow}+n}{m_{\uparrow}m_{\downarrow}+n^{2}}\frac{1}{l_{B}^{2}}+2\pi\frac{m_{\downarrow}}{m_{\uparrow}m_{\downarrow}+n^{2}}.
\label{Eq grmma}
\end{eqnarray}

As we mentioned in Sec.\ref{sec2}, $l_{B}$ is just the effective
magnetic length in the FQSH state, and $l_{B}^{2}$=$(eB)^{-1}$, B
is the effective magnetic field arising from the  strain gradient.
So the first term of Eq.(\ref{Eq grmma}) is  proportion to the
effective magnetic field B and the area. Only the second term
corresponds to the fractional statistics of the quasihole in the
spin-$\uparrow$ layer. Noting that the statistical angle
associated with moving one quasihole half-way around the other, we
finally obtain that
\begin{eqnarray}
\theta^{(\uparrow)}=\pi\frac{m_{\downarrow}}{m_{\uparrow}m_{\downarrow}+n^{2}},
\end{eqnarray}
which is accordance with the result (\ref{quasihole statistical
angles}) in Sec.\ref{sec3A}. On the other hand, if we introduce
\begin{eqnarray}
a_{x}=a_{\xi_{\uparrow}}+a_{\xi_{\uparrow}^{*}}~,~~a_{y}=i(a_{\xi_{\uparrow}}-a_{\xi_{\uparrow}^{*}}),
\label{Eq ax, ay}
\end{eqnarray}
from Eqs.(\ref{Eq a, a* E}) and (\ref{Eq ax, ay}), we have
\begin{eqnarray}
(a_{x},a_{y})&=&e\frac{m_{\downarrow}+n}{m_{\uparrow}m_{\downarrow}+n^{2}}(A_{x},A_{y})\nonumber\\
&&+\frac{m_{\downarrow}}{m_{\uparrow}m_{\downarrow}+n^{2}}\frac{1}{r^{2}}(-y,x).
\label{Eq ax, ay E}
\end{eqnarray}
One can easily see that the first term of Eq.(\ref{Eq ax, ay E})
is due to the effective magnetic field, and the second term, which
is proportion to $\frac{1}{r^{2}}(-y,x)$=$\nabla\theta$,
($\theta$=$\arg\xi_{\uparrow}$), is nothing but the `statistical'
vector potential arising  from the flux tube \cite{Arovas2,
Nayak}. Then we get the same result
\begin{eqnarray}
\theta^{(\uparrow)}=\frac{1}{2}\frac{m_{\downarrow}}{m_{\uparrow}m_{\downarrow}+n^{2}}\oint
d\mathbf{r}\cdot\nabla\theta=
\pi\frac{m_{\downarrow}}{m_{\uparrow}m_{\downarrow}+n^{2}},
\nonumber
\end{eqnarray}
as the result given in (\ref{quasihole statistical angles}).
Similarly, we can obtain the statistical angle
$\theta^{(\downarrow)}$ and the mutual statistics
$\theta^{(\uparrow\downarrow)}$ by generalized plasma analogy and
Berry phase technique, and find that they are in perfect agreement
with the results given in (\ref{quasihole statistical angles}) and
(\ref{quasiholes mutual statistics}) in Sec.\ref{sec3A}.

\section{Conclusions}
\label{sec4}

In this article, we study the topological properties of the two
types of fundamental quasihole excitation in the FQSH state by
Chern-Simons filed theory and Berry phase technique. We use the
two different approaches to calculate the fractional charge and
statistical angle of each type of quasihole excitation, as well as
the mutual statistics between the two different types of
excitation, respectively. All results obtained from the two
methods are identical.

\section*{Acknowledgement}

This work is supported by NSFC Grant No.10675108.


\begin{thebibliography}{99}

\bibitem{Klitzing}
K. v. Klitzing, G. Dorda, and M. Pepper, Phys. Rev. Lett. {\bf 45}, 494 (1980)

\bibitem{Tsui}
D. C. Tsui, H. L. Stormer, and A. C. Gossard, Phys. Rev. Lett. {\bf 48}, 1559 (1982)

\bibitem{Thouless}
D. J. Thouless, M. Kohmoto, M. P. Nightingale, and M. den Nijs, Phys. Rev. Lett. {\bf 49}, 405 (1982)

\bibitem{Wen1}
X. G. Wen and A. Zee, Phys. Rev. B {\bf 46},2290 (1992)

\bibitem{Haldane}
F. D. M. Haldane, Phys. Rev. Lett. {\bf 61}, 2015 (1988)

\bibitem{Kane1}
C. L. Kane and E. J. Mele, Phys. Rev. Lett. {\bf 95}, 226801 (2005)

\bibitem{Kane2}
C. L. Kane and E. J. Mele, Phys. Rev. Lett. {\bf 95}, 146802 (2005)

\bibitem{Zhang1}
B. A. Bernevig and S. C. Zhang, Phys. Rev. Lett. {\bf 96}, 106802 (2006)

\bibitem{Zhang2}
B. A. Bernevig, T. A. Hughes, and S. C. Zhang, Science {\bf 314}, 1757 (2006)

\bibitem{Zhang3}
M. K\"{o}nig, S. Wiedmann, C. Br\"{u}ne, A. Roth, H. Buhmann, L. W. Molenkamp, X. L. Qi, and S. C. Zhang, Science {\bf 318}, 766 (2007).

\bibitem{Levin1}
M. Levin and A. Stern, Phys. Rev. Lett. {\bf 103}, 196803 (2009)

\bibitem{Levin2}
M. Levin, F. J. Burnell, M. K. Janusz, and Ady Stern, arXiv:1108.4954

\bibitem{Qi}
X. L. Qi, Phys. Rev. Lett. {\bf 107}, 126803 (2011)

\bibitem{Wilczek1}
F. Wilczek, Fractional Statistics and Anyon Superconductivity (World Scientific, Singapore, 1990)

\bibitem{Nayak}
C. Nayak, S. H. Simon, M. Freedman, S. D. Sarma, Rev. Mod. Phys. {\bf 80}, 1083 (2008)

\bibitem{Halperin}
B. Halperin, Helv. Phys. Acta {\bf 56}, 75 (1983)

\bibitem{Wen2}
X. G. Wen and A. Zee, Phys. Rev. Lett. {\bf 69}, 1811 (1992)

\bibitem{Zhang4}
S. C. Zhang, H. Hansson, and S. Kivelson, Phys. Rev. Lett. {\bf 62}, 82 (1989)

\bibitem{Wilczek2}
F. Wilczek and A. Zee, Phys. Rev. Lett. {\bf 51}, 2250 (1983)

\bibitem{Polyakov}
A. M. Polyakov, Mod. Phys. Lett. {\bf A3}, 325 (1988)

\bibitem{Arovas1}
D. Arovas, J. R. Schrieffer, and F. Wilczek, Phys. Rev. Lett. {\bf 53}, 722 (1984)

\bibitem{Laughlin}
R. B. Laughlin, Phys. Rev. Lett. 50, 1395 (1983)

\bibitem{Forrester}
P. J. Forrester and B. Jancovici, J. Phys. (Pairs) Lett. {\bf 45}, L583 (1984)

\bibitem{Arovas2}
D. Arovas, J. R. Schrieffer, F. Wilczek, and A. Zee, Nucl. Phys. B
{\bf 251}, 117 (1985)

\end{thebibliography}
\end{document}